\documentclass[usenatbib]{mn2e}%MN-13-3132-MJ submitted 3/12/13
\usepackage{natbib}
\usepackage{epsfig}

\catcode`\@=11
\def\gta{\ifmmode{\,\mathrel{\mathpalette\@versim>\,}}
    \else{$\,\mathrel{\mathpalette\@versim>}\,$}\fi}
\def\lta{\ifmmode{\,\mathrel{\mathpalette\@versim<\,}}
    \else{$\,\mathrel{\mathpalette\@versim<}\,$}\fi}
\def\@versim#1#2{\lower 2.9truept \vbox{\baselineskip 0pt \lineskip
    0.5truept \ialign{$\m@th#1\hfil##\hfil$\crcr#2\crcr\sim\crcr}}}
\catcode`\@=12  
\renewcommand{\[}{\begin{equation}}
\renewcommand{\]}{\end{equation}}
\let\boldgrk=\gkvecten
\let\boldgrksc=\gkvecseven

\def\gkthing#1{{\mathchoice%
	{\hbox{{\boldgrk\char#1}}}
	{\hbox{{\boldgrk\char#1}}}
	{\hbox{{\boldgrksc\char#1}}}
	{\hbox{{\boldgrksc\char#1}}}}}
\def\lamR{\lambda_{\rm R}}
\def\Ve{v_{\rm e}}

\def\valpha{\gkthing{11}}

\def\vtheta{\gkthing{18}}

{\newif\ifnotend
\notendtrue
\def\veclist{ABCDEFGHIJKLMNOPQRSTUVWXYZabcdefghijklmnopqrstuvwxyz.}
\def\top#1#2.{#1}
\def\tail#1#2.{#2.}
\loop\expandafter\xdef\csname v\expandafter\top\veclist\endcsname%
{{\noexpand\bf\expandafter\top\veclist}}
\edef\veclist{\expandafter\tail\veclist}
\if\veclist.\notendfalse\fi\ifnotend\repeat}
{\newif\ifnotend
\notendtrue
\def\veclist{ABCDEFGHIJKLMNOPQRSTUVWXYZ.}
\def\top#1#2.{#1}
\def\tail#1#2.{#2.}
\loop\expandafter\xdef\csname c\expandafter\top\veclist\endcsname%
{{\noexpand\cal\expandafter\top\veclist}}
\edef\veclist{\expandafter\tail\veclist}
\if\veclist.\notendfalse\fi\ifnotend\repeat}

\def\df{{\sc df}}

\def\fracj#1#2{{\textstyle{#1\over#2}}}
\def\pa{\partial}
\def\ex#1{\left\langle#1\right\rangle}

\def\kpc{\,{\rm kpc}}

\def\e{{\rm e}}
\def\d{{\rm d}}

\def\Enc{E_{\rm rand}}
\def\figref#1{Fig.~\ref{#1}}
\newcommand{\beq}{\begin{equation}}
\newcommand{\eeq}{\end{equation}}

\title[Self-consistent flattened isochrone models]
{Self-consistent flattened isochrones}

\author[James Binney]{James  Binney\thanks{E-mail:
binney@thphys.ox.ac.uk}\\
Rudolf Peierls Centre for Theoretical Physics, Keble Road, Oxford OX1 3NP, UK\\
}

\begin{document}

\date{Draft, February 7, 2013}

\pagerange{\pageref{firstpage}--\pageref{lastpage}} \pubyear{2012}

\maketitle

\label{firstpage}

\begin{abstract}
We present a family of self-consistent axisymmetric stellar systems that have
analytic distribution functions (\df s) of the form $f(\vJ)$, so they depend
on three integrals of motion and have triaxial velocity ellipsoids. The
models, which are generalisations of H\'enon's isochrone sphere, have four
dimensionless parameters, two determining the part of \df\ that is even in
$L_z$, and two determining the odd part of the \df\ (which determines the
azimuthal velocity distribution). Outside their cores, the velocity ellipsoids
of all models tend to point to the model's centre, and we argue that this
behaviour is generic, so near the symmetry axis of a flattened model, the
long axis of the velocity ellipsoid is naturally aligned with the symmetry
axis and not perpendicular to it as in many published dynamical models of
well-studied galaxies.  By varying one of the \df's parameters, the intensity
of rotation can be increased from zero up to a maximum value set by the
requirement that the \df\ be non-negative. Since angle-action coordinates are
easily computed for these models, they are ideally suited for perturbative
treatments and stability analysis.  They can also be used to choose initial
conditions for an N-body model that starts in perfect equilibrium and to
model observations of early-type galaxies.  The modelling technique
introduced here is readily extended to different radial density profiles,
more complex kinematics, and multi-component systems. A number of important
technical issues surrounding the determination of the models' observable
properties are explained in two appendices.
\end{abstract}


\section{Introduction} Although real galaxies are by no means in states of
dynamical equilibrium, equilibrium models have a fundamental role to play in
the study of galaxies and the Universe. One reason for this is that most of a
galaxy's mass is thought to reside in dark matter that can only be traced by its
gravitational field, and we can map this field
through the kinematics of field stars only to the extent that the stars are
in dynamical equilibrium. A further reason is that modest deviations of
galaxies from dynamical equilibrium are best modelled by perturbing
equilibrium models -- most successful branches of physics, from celestial
mechanics to high-energy physics through plasma physics and quantum
condensed-matter physics, comprise applications of perturbation theory.

Currently the range of equilibrium galaxy models is extraordinarily limited.
The theory of globular clusters rests to a great extent on the equilibrium
models introduced by \cite{Michie} and popularised by \cite{King}. Our still
very limited understanding of spiral structure owes much to the
two-dimensional equilibrium models of \cite{Kalnajs}, Zang \& Toomre
\citep{Toomre} and Evans \citep{Evans94,ReadEvans,EvansSellwood}. 

On account of the paucity of fully three-dimensional equilibrium models,
several recent analyses of our Galaxy's halo have relied on  models which
have
distribution functions (\df s) of the form 
 \[\label{eq:beta}
f(E,L)=L^{-2\beta}F(E),
\]
 where $E$ is a star's energy and $L$ is the magnitude of its angular
momentum \citep{Deasonxx}. Models with a \df\ of this form only make
dynamical sense to the extent that the gravitational potential can be
considered to be spherically symmetric, which it probably cannot be in the
solar neighbourhood. Moreover, in the case of radial bias, the \df\
(\ref{eq:beta}) implies infinite phase-space density in the limit $L\to0$ of
radial orbits, which is inherently implausible and potentially compromises
the model's stability \citep{FridP,PalmerP}, while in the case $\beta<0$ of
tangential bias, the \df\ implies a distribution of velocities that is
physically implausible because it is bimodal in $v_\phi$ \citep{FermaniSa}.
Hence \df s of the form (\ref{eq:beta}) do not constitute a satisfactory
basis for dynamical modelling.

Although galaxies are probably never precisely axisymmetric, they are often
sufficiently nearly so for axisymmetric models to be valuable starting
points from which better-fitting models may be derived by perturbation
theory. In this paper we present a new class of axisymmetric models. In a
subsequent paper it will be shown how the technique we introduce here can be
used to produce a wide range of axisymmetric models.

Already in the paper in which Jeans introduced his theorem, it was clear that
the equilibrium \df\ of our Galaxy cannot be a function of the form
$f(E,L_z)$ that depends only on the classical integrals in an axisymmetric
potential. Dependence on a third integral $I_3$ is essential, and the field
has been held back by the want of an analytic formula for $I_3(\vx,\vv)$. In
the 1970s it became evident that the \df s of elliptical galaxies also
involve $I_3$ in an essential way \citep{Bertola,Binney76,Daviesetal},
although recent systematic surveys have cast a new light on this conclusion
\citep{Cappellarietal,Emsellem}. Nevertheless, it remains true that an abundance of
observational material indicates that realistic equilibrium models of real
galaxies must depend on $I_3$ in addition to $E$ and $L_z$.

Numerical orbit integrations in the 1960s showed that generic orbits in
flattened axisymmetric potentials respect a third integral $I_3$
\citep{HenonH,Ollongren}. But these experiments did not lead to useful
analytic expressions for $I_3(\vx,\vv)$. A promising attack on this problem
through Hamiltonian perturbation theory was pursued by Contopoulos and his
collaborators \citep{Contopoulos}, but this line of attack was frustrated by
the fact that the vertical oscillations of most stars are far from harmonic,
so their orbits are not readily treated as perturbed harmonic oscillators.
Moreover the coupling between a star's radial and vertical oscillations is
fundamental to their dynamics, so neither motion should be considered in
isolation.

\cite{Eddington} and \cite{Stackel} showed that analytic expressions for
$I_3$ can be obtained for a certain class of potentials that are now known as
St\"ackel potentials.  \cite{deZeeuw} showed that many of these potentials
are generated by remarkably galaxy-like density distributions, and he
clarified the nature of orbits in these potentials. The present paper relies
on a technique, the ``St\"ackel Fudge'' \citep[][hereafter B12a]{Binney12a},
which is an extension of this classic work. This approximation consists of
applying to an arbitrary gravitational potential $\Phi(R,z)$ formulae that
would be valid if the potential were of St\"ackel's form even though the
potential does not have this form.

\begin{figure}
\centerline{\epsfig{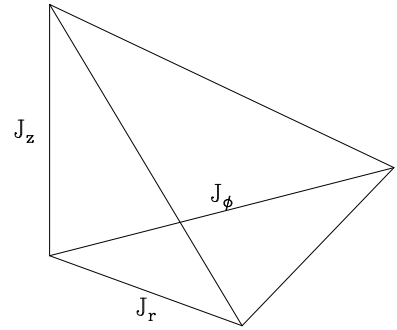}}
\caption{The tetrahedral zone in action space occupied by orbits with
energies less than some maximum value.}
\label{fig:Jspace}
\end{figure}

Since any function of a star's isolating integrals is itself an isolating
integral, there is in principle  enormous freedom in the choice of arguments
of a galaxy's \df. However, certain integrals stand out for special
consideration: the action integrals. These alone can be embedded in a
canonical coordinate system, and their conjugate (angle) variables have the
remarkable property of increasing linearly in time:
 \[\label{eq:theta_of_t}
\theta_i(t)=\theta_i(0)+\Omega_i t.
\]
 Action integrals are unique up to a set of discrete canonical
transformations that map between rational linear combinations of any given
set of actions, and map the angle variables into integer linear combinations
of the given angles. Hence when combined with reasonably physical
requirements such as ``the radial action $J_r$ should quantify the extent of
radial oscillations'' and ``the vertical action $J_z$ should quantify the
extent of vertical oscillations'', actions are uniquely defined. This
uniqueness greatly facilitates the comparison of models by making it possible
to compare the density with which orbits in slightly different potentials are
populated.  Therefore it is natural to require the \df\ to be a function
$f(\vJ)$ of the actions.

To derive from $f(\vJ)$ the observable properties of a model, we need
expressions for the actions as function $\vJ(\vx,\vv)$ of the ordinary
phase-space coordinates. Here we use the St\"ackel Fudge from B12a.  Numerical
experiments presented in B12a show that the actions and angles that one
obtains from the Fudge are significantly more accurate than those obtained
from the adiabatic approximation \citep{B10}, and they are also valid for
orbits that move far from the equatorial plane and thus lie outside the range
of validity of the adiabatic approximation.  Whereas \cite{Binney12b} and
\cite{Binneyetal13} only required actions for orbits that pass within
$\sim2\kpc$ of the solar neighbourhood, we require actions throughout our
models. This change has prompted us to undertake significant revisions of the
scheme described by B12a to obtain them with increased accuracy and reduced
numerical effort. These revisions are described in Appendix I.

In Section \ref{sec:DF} we define our \df s, which are derived from the \df\
of the isochrone sphere. In Section \ref{sec:relax} we explain how a model
is relaxed to its self-consistent configuration and detail checks of
the numerical accuracy.  In Section \ref{sec:obs} we describe the observable
properties of some specific models and refine our choice of \df. In Section
\ref{sec:appls} we discuss some potential applications of these models,
including choice of initial conditions for N-body models, studies of symmetry
breaking in stellar systems and related perturbation analyses, and modelling
observations of early-type galaxies. Section \ref{sec:conclude} sums up. Two
Appendices give numerical details that are likely to be useful when building
the models.

\section{The distribution functions}\label{sec:DF}

The isochrone potential \citep{HenonI}
 \[
\Phi_{\rm I}(r)=-{GM\over{b+\sqrt{r^2+b^2}}},
\]
 where $M$ is the model's total mass and $b$ is its scale length, is highly
unusual in that it admits analytic expressions for the both angles and actions
as functions of $(\vx,\vv)$ \citep[e.g.][\S3.5]{BT08}. Moreover, the
associated Hamiltonian
\[
H_{\rm I}(r,v)=\fracj12v^2+\Phi_{\rm I}(r)
\]
 can be written as a simple function of the actions
\[\label{eq:HJ}
H_{\rm I}(\vJ)=-{(GM)^2\over2[J_r+\frac12(L+\sqrt{L^2+4GMb}\,)]^2 },
\]
 where $L\equiv |J_\phi|+J_z$ is the total angular momentum.
The \df\ $f_{\rm I}(H)$ that self-consistently generates $\Phi_{\rm I}$ can
be derived from Eddington's inversion formula. It proves to be
\citep{HenonIII}
\begin{eqnarray}\label{eq:fH}
f_{\rm I}(H_{\rm I})&=&{1\over\sqrt{2}(2\pi)^3(GMb)^{3/2}}{\surd
\cH\over[2(1-\cH)]^4}\Bigg[27-66\cH+320\cH^2\nonumber\\
&&\hskip-7mm-240\cH^3+ 64\cH^4+3(16\cH^2+28\cH-9)
{\sin^{-1}\surd\cH\over\sqrt{\cH(1-\cH)}}\Bigg],
\end{eqnarray}
 where
 \[
\cH=-{H_{\rm I}b\over GM}
\]
 is the dimensionless relative energy. We obtain $f_{\rm I}(\vJ)$ by using equation
(\ref{eq:HJ}) to eliminate $\cH$ from equation \ref{eq:fH}.

$f_{\rm I}$ generates a spherical model because $J_\phi$ and $J_z$ appear in
it on an equal footing. We can flatten the model by causing the \df\ to
decrease with increasing $J_z$ faster than with increasing $J_r$ or $J_\phi$.
A \df\ that achieves this goal is
\[\label{eq:foJ}
f_{\valpha}(\vJ)\equiv (\alpha_r\alpha_\phi\alpha_z)f_{\rm I}(\alpha_rJ_r,\alpha_\phi
J_\phi,\alpha_z J_z)
\]
 where $\valpha$ is a triple of constants with
$\alpha_z>\max(\alpha_r,\alpha_\phi)$.

The radial density profile of a spherical model is largely determined by $\d
N/\d E$, the number of stars per unit energy at $E$ \citep[][Fig.~4.5 and
\S4.4]{BT08}.  That is, the radial density profile of a model changes only
modestly when stars are shifted over a surface of constant $E$.  A
tangentially anisotropic model is created if stars are shifted over the
surface towards the line $|J_\phi|+J_z=L_c(E)$, where $L_c(E)$ is the energy
of a circular orbit of energy $E$ (\figref{fig:Jspace}). Conversely, a radially anisotropic model
is created if stars are moved over surfaces of constant $E$ towards the $J_r$
axis. Given that the \df\ is invariably a decreasing function of all three
actions, when we replace $J_i$ in $f$ with $\alpha_iJ_i$, the value
of $f$ is decreased for a given value of $\vJ$ if $\alpha_i>1$, and increased
otherwise. We would like these changes to average to zero over a surface of
constant $E$ so we can interpret them as the results of moving stars over
these surfaces while keeping $\d N/\d E$ unchanged. For $\alpha_i\simeq1$,
the change in $f$ caused by the substitution $J_i\to\alpha_iJ_i$ is
 \[
\delta f\simeq {\d f\over\d H_{\rm I}}\sum_i{\pa H_{\rm I}\over\pa
J_i}(\alpha_i-1).
\]
 In order to preserve the radial density profile, we want the integral of
$\delta f$ over a surface of constant $E$ to vanish. This will be
approximately the case if
 \[\label{eq:dfzero}
\sum_i\Omega_i(\overline{\vJ})(\alpha_i-1)=0,
\]
 where $\overline{\vJ}$ is the barycentre of the surface of constant $E$,
i.e., the action of the form
$\overline{\vJ}=(\overline{J},\overline{J},\overline{J})$ that lies in the
surface. For the isochrone potential one finds
 \[\label{eq:defsJbar}
\overline{J}=\fracj13\left({2GM\over\sqrt{-2E}}-\sqrt{{(GM)^2\over-2E}-3GMb}\,\right).
\]

So long as we require equation (\ref{eq:dfzero}) to be satisfied,
only two of the scaling factors $\alpha_i$ should be considered independent
for the third can be obtained from this equation. Below we explore models in
which $\alpha_\phi$ and $\alpha_z$ are taken to be constants and $\alpha_r$
becomes through equation (\ref{eq:dfzero}) a function of energy:
\[\label{eq:defsa1}
\alpha_r=\alpha_{r0}(\overline{\vJ})\equiv1-{\Omega_L\over\Omega_r}(\alpha_\phi+\alpha_z-2),
\]
 where $\Omega_L(\overline{\vJ})$ and $\Omega_r(\overline{\vJ})$ are the
angular and radial frequencies of the specified orbit in the isochrone sphere
defined by equation (\ref{eq:theta_of_t}). 

\subsection{Rotating models}

The \df\ of the isochrone is an even function of the angular momentum
$L_z=J_\phi$ and this property is preserved by the transformation
(\ref{eq:foJ}) proposed above. When the \df\ is an even function of $J_\phi$, the
model does not rotate. It can be set rotating by adding to the given even
\df, $f_+(\vJ)$, a \df\ $f_-(\vJ)$ that is an odd function of $J_\phi$. Then the
complete \df\ becomes
 \[\label{eq:f_of_k}
f(\vJ)=(1-k)f_+(\vJ)+kf_-(\vJ),
\]
 where $k$ is a free parameter that allows one to vary the
rotation speed from zero at $k=0$ up to a maximum value that is set by the
requirement that $f$ is never negative.

Given an even \df\ $f_+(\vJ)$, a natural definition of
an odd \df\ is
 \[
f_-(\vJ)=g(J_\phi)f_+(\vJ),
\]
 where $g(J_\phi)\le1$ is an odd function of $J_\phi$. A maximally rotating
model is obtained by choosing $k=\frac12$ and $g(x)=\hbox{sign}(x)$, but with
this choice of $g$, $f$ is discontinuous on the plane $J_\phi=0$, where its
absolute value is typically large. To avoid such a discontinuity we adopt
\[
g(J_\phi)=\tanh\left({\chi J_\phi\over\sqrt{GMb}}\right).
\]
 Here $\chi$ is a dimensionless parameter that specifies the steepness of the
rotation curve near the origin: the larger the value of
$\chi$, the more steeply the curve rises. With these choices we have 
 \begin{eqnarray}
\overline{v}_\phi(\vx)&=&{k\over1-k}
{\int_0^\infty\d v_\phi\,v_\phi g(Rv_\phi)\int\d v_r\d v_z f_+(\vx,\vv)
\over\int_0^\infty\d v_\phi\int\d v_r\d v_z f_+(\vx,\vv)}\nonumber\\
\overline{v_\phi^2}(\vx)&=&{\int_0^\infty\d v_\phi\,v_\phi^2\int\d v_r\d v_z f_+(\vx,\vv)
\over\int_0^\infty\d v_\phi\int\d v_r\d v_z f_+(\vx,\vv)}.
\end{eqnarray}
 Naturally the azimuthal velocity dispersion is 
\[
\sigma_\phi^2(\vx)=\overline{v_\phi^2}-(\overline{v}_\phi)^2.
\]

\begin{figure}
\centerline{\epsfig{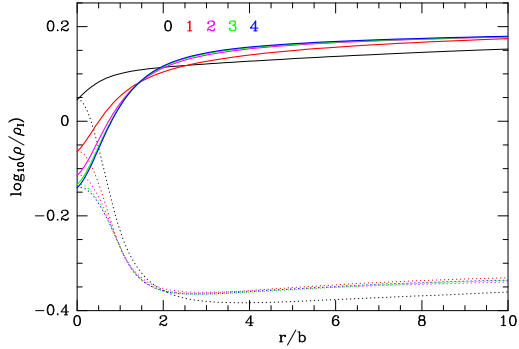}}
\caption{Convergence of the model with $\alpha_\phi=0.7$ and
$\alpha_z=1.4$. We plot  $\log_{10}[\rho(r,\theta)/\rho_{\rm
I}(r)]$, where $\rho_{\rm I}(r)$  is the density of the isochrone sphere, for
two values of $\theta$: full curves are for a ray that lies close to the
major axis while dotted curves are for a ray that lies close to the minor
axis. The colour of the curves indicates which iterate of the potential was
used for the density evaluation.}
\label{fig:0itr}
\end{figure}

\section{Finding the self-consistent potential}\label{sec:relax}

Given values for $\alpha_\phi,\alpha_z,k$ and $\chi$ to specify a \df, one
has to recover the corresponding model's density and self-consistent
potential from iterations.  One adopts some trial gravitational potential
$\Phi_0(R,z)$ and computes the density implied by $\Phi_0$ and the \df\ on a
grid in the $(R,z)$ plane. Then one computes the potential $\Phi_{1/2}$
implied by this density distribution, and computes the extrapolated potential
 \[
\Phi_1=(1+\gamma)\Phi_{1/2}-\gamma\Phi_0.
\]
 Now one repeats  this cycle with $\Phi_0$ replaced by $\Phi_1$. A 
positive value of $\gamma$ speeds convergence of these iterations; if
$\gamma$ is above a threshold, numerical instability sets in. For the \df s
explored here, $\gamma=0.5$ works well.

\begin{figure}
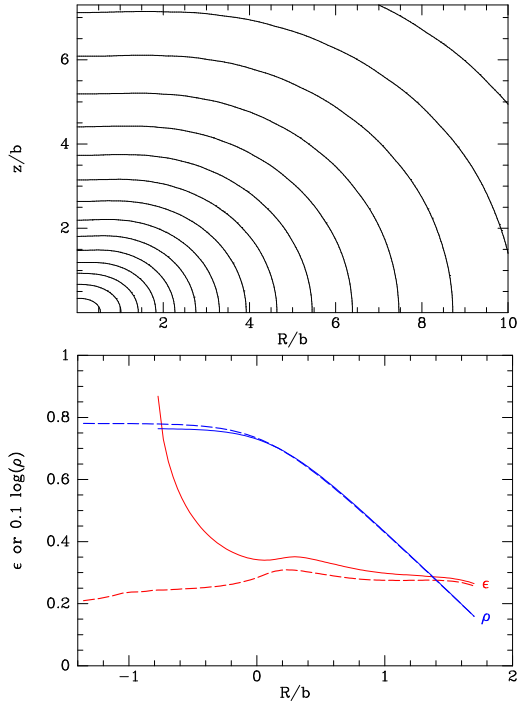

\centerline{\epsfig{file=Hiso-0.30_0.40cont.ps,width=.8\hsize}}
\centerline{\epsfig{file=eps-0.3_0.4.ps,width=.8\hsize}}
 \caption{Top: contours of equal density in the meridional plane for the
model with $\alpha_\phi=0.7$ and $\alpha_z=1.4$. Below: ellipticity
$\epsilon=1-c/a$ as a function of radius in this model (full red) and $0.1$
times the logarithm to base 10 of the density along the model's major axis
(full blue). The broken red and blue curves show the
ellipticity and density when the model is modified by using equation
(\ref{eq:defsai}) with $\alpha_{\phi0}=0.7$ to ensure that
$\sigma_R^2-\overline{v_\phi^2}$ tends to zero at the origin faster than
$R$.} \label{fig:0cont}
\end{figure}

Figs~\ref{fig:0itr} and \ref{fig:0cont} show results obtained with
$\alpha_\phi=0.7$, $\alpha_z=1.4$ when the trial potential $\Phi_0$ is that of an
isochrone sphere flattened to axis ratio $0.7$.  In \figref{fig:0itr} a black
dotted curve shows the density on the minor axis yielded by the \df\ in the
trial potential divided by the density of the corresponding isochrone sphere.
The red curve shows the corresponding ratios after one adjustment to the
potential, followed by the magenta, green and blue curves for the second through
the fourth adjustments of the potential. The near
coincidence of the green and blue curves demonstrates that the iterations
have essentially reached a stationary point. The full curves in
\figref{fig:0itr} show the corresponding results for the major axis. Where
the blue curves run nearly horizontally, the density profile is essentially
proportional to that of the underlying isochrone sphere, as planned. Within
$r\simeq2b$ the full and dotted blue curves necessarily converge on a point,
which indicates the ratio of the central density in the final model to that
of the isochrone sphere. 

Finding the density and potential that correspond to a given \df\ using five
iterations of the density takes $\sim3$ CPU hours on a desk-top machine. If
the key loop of the code is parallelised a model can be constructed in less
than half an hour.

\subsection{Checks of accuracy}

Many checks on the accuracy of the computations are possible. First, one can
compute orbits in a model's potential and evaluate the actions at
different points along the orbit. The fluctuations in the computed actions
are then typically $10^{-5}$ of $J_r+J_z$. Errors in the evaluation of forces
and the potential by interpolation on the radial grid (see Appendix II) could alone account for errors of this magnitude.
Interpolation of the grid in action space (Appendix I) introduces
errors of order $0.2\%$ in the actions used for the evaluation of moments. 

When the apparatus is used to evaluate the density of a model that is
essentially the isochrone sphere by setting $\alpha_i=1$ in the \df\ and
adopting the potential of an isochrone sphere that has been squashed to axis
ratio $q=0.999$ (the case $q=1$ gives rise to a singularity in the equations
employed), the density of the isochrone sphere is recovered to parts in
$10^4$.

The integral $\int\d^3\vJ\,f(\vJ)$ yields the mass of the model divided by
$(2\pi)^3$ and from the definition (\ref{eq:foJ}) of the \df\ it follows that
the model's mass is the same as that of the isochrone, $M$.  Numerically we
can compute the mass of a model that lies within radius $r$ by computing
$r^2/G$ times the monopole component of the radial component of the
gravitational force $F_r(r)$. In the typical case of the flattened model
plotted in \figref{fig:0itr}, with the outer edge of the grid set to $50b$,
the radial force implies that $0.9552$ lies inside $50b$, while in the
isochrone sphere $0.9606M$ lies inside $50b$. The discrepancy between these
two masses is a small fraction of the mass that lies outside the grid in the
spherical case. This finding is consistent with there being no error in the
computed mass and potential. Note too that we expect to obtain the same mass
by integrating $f$ over $\vJ$ as we do by integrating $\rho$ over $\vv$ only
because the $(\vtheta,\vJ)$ system is canonical. Hence the mass check tests
the validity of the St\"ackel Fudge.
 
The virial theorem provides three useful checks on both the validity of the
calculations and the convergence of the potential. For an axisymmetric system
the tensor virial theorem has two non-trivial components, one,
$2K_{RR}=-W_{RR}$ associated with the cylindrical radius $R$ and a vertical
component, $2K_{zz}=-W_{zz}$. The sum of these two components constitutes the
scalar virial theorem, $2K=-W$. After five iterations one finds in a typical
case that $2+W/K\simeq-2.9\times10^{-3}$ while
$2+W_{RR}/K_{RR}\simeq1.5\times10^{-3}$ and
$2+W_{zz}/K_{zz}\simeq1.05\times10^{-3}$.

A further check on accuracy is provided by the Jeans equations. As discussed
below, these are satisfied to within the precision with which we are 
evaluating spatial derivatives within the final model.

\begin{figure}
\centerline{\epsfig{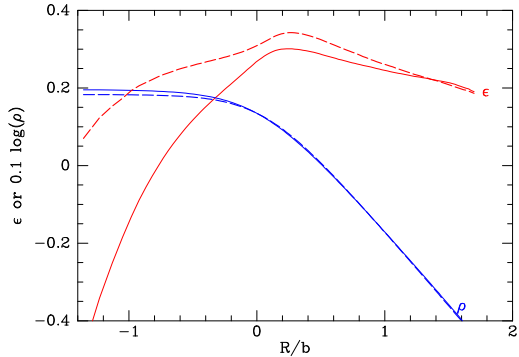}}
 \caption{Full red: ellipticity $\epsilon=1-c/a$ as a function of radius for the
model with $\alpha_\phi=1$ and $\alpha_z=1.5$. Full blue: 0.1 times the
logarithm to
base 10 of the density along this model's major axis. Broken red and broken blue:
ellipticity and density when the  model is modified 
by using equation
(\ref{eq:defsai}) with $\alpha_{\phi0}=1$ to ensure that
$\sigma_R^2-\overline{v_\phi^2}$ tends to zero at the origin faster than
$R$.} \label{fig:2eps}
\end{figure}

\section{Observables}\label{sec:obs}

We consider first non-rotating models. The half-mass radius of the isochrone
sphere is $3.06b$.

\subsection{Ellipticity profiles}

The top panel of \figref{fig:0cont} shows contours of constant density of the
model with $\alpha_\phi=0.7$, $\alpha_z=1.4$ in the $(R,z)$
plane, with four contours per decade. The flattening of the model is evident.
The black curve in the lower panel of \figref{fig:0cont} shows the
ellipticity $\epsilon\equiv1-c/a$ of isodensity surfaces as a function of the
length $a$ of the semi-major axis.  From $r\simeq0.7b$ to $r=30b$ $\epsilon$
is nearly constant, falling from $0.35$ to $0.3$. Inside $r=b$, the
ellipticity $\epsilon$ rises steeply. In fact in this model the peak density
is not reached at the centre but in the equatorial plane at $R=0.089b$. The
density at the centre is, however, $0.9926$ times the peak density, so the
upward lurch of the ellipticity curve in \figref{fig:0cont} is caused by a
very minor depression in the central density.

The black curve in \figref{fig:2eps} shows the ellipticity as a function of
the logarithm of radius in the model obtained by setting $\alpha_\phi=1$ and
$\alpha_z=1.5$.  The choice $\alpha_\phi=1$ implies that the dependence of
the \df\ on $J_\phi$ is precisely that of the isochrone sphere, so equation
(\ref{eq:dfzero}) now causes $\alpha_r$ to be materially smaller than unity,
so the \df\ becomes radially biased.  The ellipticity peaks at $\epsilon=0.3$
just outside the core and further out slowly declines, reaching
$\epsilon=0.24$ at $r=10b$. Inside the core $\epsilon$ plunges to negative
values, passing zero at $r=0.17b$ and reaching $\epsilon=-0.4$ $r=0.05b$.
Thus the innermost part of the model is prolate rather than oblate. It is
instructive to understand why.

\begin{figure}
\centerline{\epsfig{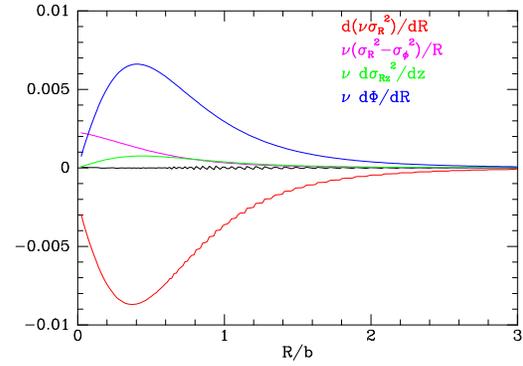}}
\caption{The values of the terms in the radial Jeans equation for the prolate
model $\alpha_\phi=1,\alpha_z=1.5$. The black curve shows the sum of these
terms and would ideally be everywhere precisely zero.}\label{fig:Jeans}
\end{figure}

\begin{figure}
\centerline{\epsfig{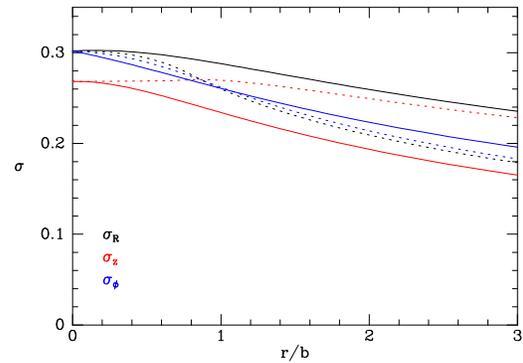}}
\caption{The radial (black), azimuthal (blue) and vertical (red) components
of velocity dispersion in the radially biased, non-rotating model $\alpha_\phi=1,\alpha_z=1.5$.
Full curves show values in the equatorial plane and dashed curves those along
the symmetry axis. The unit of velocity is $\sqrt{GM/b}$.}\label{fig:sig}
\end{figure}

In \figref{fig:Jeans} we plot the terms that appear in the radial Jeans
equation \citep[][eq 4.222a]{BT08}. When we evaluate this equation in the
equatorial plane, where $\sigma^2_{Rz}=0$ by symmetry, it becomes
 \[
{\pa(\nu\sigma_R^2)\over\pa R}+\nu\left({\pa\sigma^2_{Rv}\over\pa z}
+{\sigma_R^2-\overline{v_\phi^2}\over R}+{\pa\Phi\over\pa R}\right)=0.
\]
 The figure shows that the dominant terms in this equation are the first and
last terms, which are plotted in red and blue, respectively. The other two
terms are comparable outside $r=0.6b$ but further in the anisotropy term
(that involving $\sigma_R^2-\overline{v_\phi^2}$), shown in magenta, first
dominates the term involving $\sigma^2_{Rz}$ and then dominates the
radial-force term, which has to vanish at the origin. It acts in the same
sense as the radial-force term, i.e., inwards, and the pressure term (plotted
in red) has to
grow in magnitude to counteract it. It is this term that makes the model's
centre prolate. \figref{fig:sig} shows the radial dependencies of $\sigma_R$,
$\sigma_\phi$ and $\sigma_z$ along this model's principal axes. We see that
at the centre is approached through the equatorial plane, $\sigma_\phi$,
plotted in blue, does approach $<\sigma_R$ (black) from below, but
\figref{fig:Jeans} shows that it does not do so quite fast enough to prevent
the anisotropy term in the radial Jeans equation growing as the centre is
approached.

An analogous analysis of the Jeans equations for the oblate model plotted in
\figref{fig:0cont} shows that in this model the anisotropy term is slightly
negative, so it pushes material away from the centre and thus gives rise to
the slight central depression in the density that is responsible to the
central spike in the model's ellipticity curve.

Physically the steep rise in the ellipticity of the tangentially biased model
and the central prolateness of the radially biased model are not very
significant because in the nearly homogeneous core a small difference between
the densities distance $r$ from the origin along the minor and major axes can
translate into a large ellipticity of the isodensity surfaces. Nonetheless,
these central ellipticity changes detract from the models' elegance, and we
seek a modification of the \df\ that will moderate or eliminate them.

From the Jeans equations it is clear that the key is to ensure that
$\sigma_R\to\sigma_\phi$ fast enough as $R\to0$. Since the \df\ of the
isochrone sphere has $\sigma_R=\sigma_\phi$ everywhere, our goal should be
attainable by ensuring that $\alpha_R\to\alpha_\phi$ as $|\vJ|\to0$. Setting
$\alpha_R=\alpha_\phi\equiv\alpha_0$ in equation (\ref{eq:dfzero}) we find
that
 \[
\alpha_0(\overline{\vJ})=1-{\Omega_L\over\Omega_L+\Omega_r}(\alpha_z-1).
\]
 with $\overline{J}$ given by equation (\ref{eq:defsJbar}). We can now require
that both $\alpha_r$ and $\alpha_\phi$ tend to $\alpha_0$ as $|\vJ|\to0$ by
writing
 \begin{eqnarray}\label{eq:defsai}
\alpha_r(\overline{\vJ})&=&(1-\psi)\alpha_0+\psi\alpha_{r0}\nonumber\\
\alpha_\phi(\overline{\vJ})&=&(1-\psi)\alpha_0+\psi\alpha_{\phi0},
\end{eqnarray}
 where $\alpha_{\phi0}$ is a given constant, $\alpha_{r0}$ is computed from
equation with $\alpha_\phi$ replaced by $\alpha_{\phi0}$, and
 \[
\psi(\overline{\vJ})\equiv\tanh(\overline{J}/\sqrt{GMb})
\]
 is a function that vanishes at the origin and is essentially unity for
values of the argument bigger than $\sim\sqrt{GMb}$.

The broken red curve in the lower panel of Fig.~\ref{fig:0cont} shows the run
of ellipticity we obtain when we use equations (\ref{eq:defsai}) with
$\alpha_{\phi0}=0.7$ and $\alpha_z=1.4$. The steep central rise in $\epsilon$
of the original model has been replaced by a modest decline to a central
value just above 0.2. The dashed red curve in \figref{fig:2eps} shows the
ellipticity of the radially biased model when equation (\ref{eq:defsai}) is
employed.  The model now becomes slightly oblate rather than prolate deep in
the core.

\begin{figure*}
\centerline{ \epsfig{file=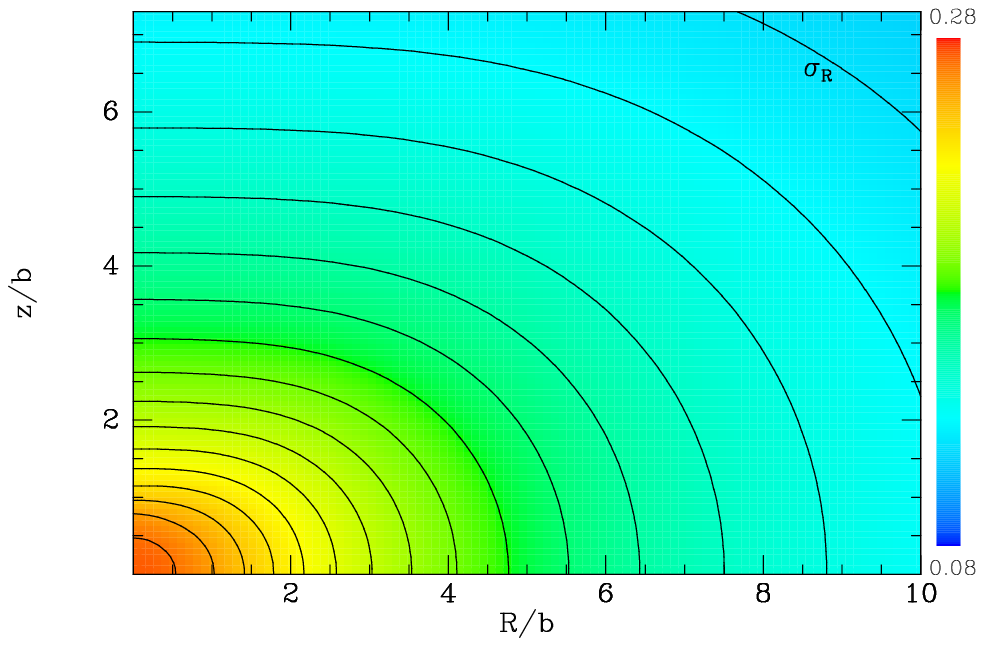,width=.32\hsize}
\epsfig{file=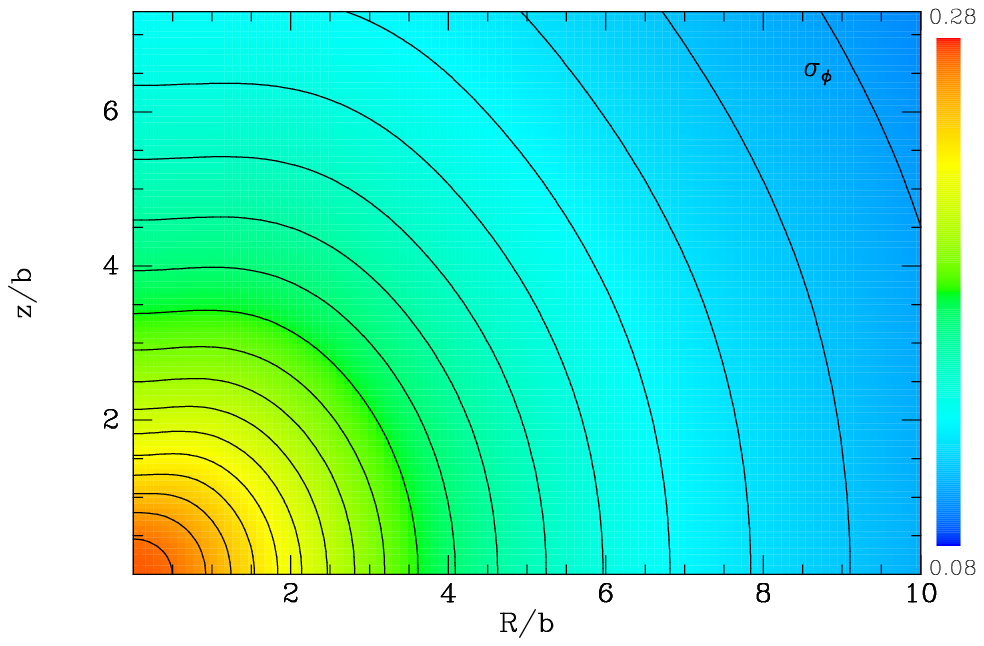,width=.32\hsize}
\epsfig{file=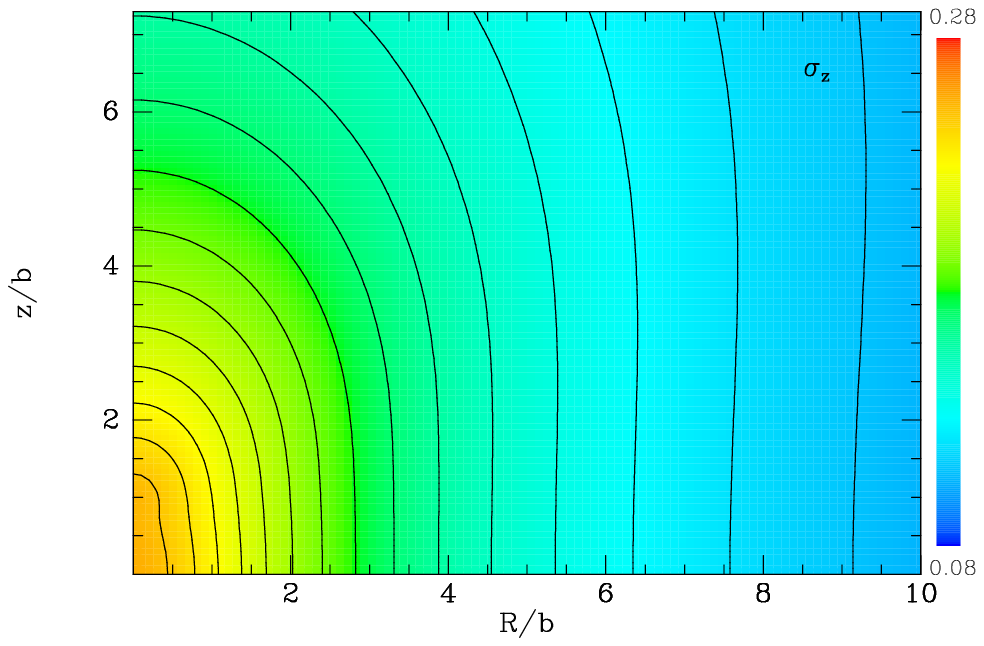,width=.32\hsize}}
 \caption{Magnitudes of
the radial (left), azimuthal (centre) and vertical (right) components of
velocity dispersion for the tangentially biased, non-rotating model with
$\alpha_{\phi0}=0.7$ and $\alpha_z=1.4$. The unit of velocity is
$\sqrt{GM/b}$.} \label{fig:1srpz}
\end{figure*}

\begin{figure*}
\centerline{
\epsfig{file=F1iso-0.00_0.50sr.ps,width=.32\hsize}
\epsfig{file=F1iso-0.00_0.50sphi.ps,width=.32\hsize}
\epsfig{file=F1iso-0.00_0.50sz.ps,width=.32\hsize}}
\caption{Magnitudes of the radial (left), aziuthal (centre) and vertical
(right) components of velocity dispersion
for the radially biased, non-rotating model with $\alpha_{\phi0}=1$ and
$\alpha_z=1.5$.  The unit of velocity is $\sqrt{GM/b}$.}
\label{fig:2srpz}
\end{figure*}

\begin{figure}
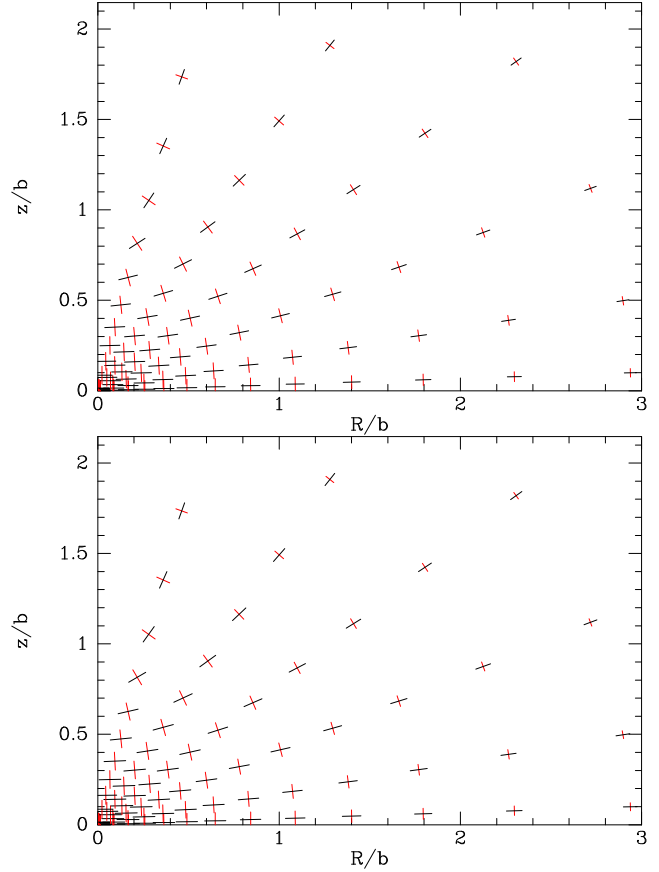

\centerline{\epsfig{file=F1iso-0.30_0.40cross.ps,width=\hsize}}
\centerline{\epsfig{file=F1iso-0.00_0.50cross.ps,width=\hsize}}
\caption{Orientation of the velocity ellipsoids for the azimuthally biased,
non-rotating model with $\alpha_{\phi0}=0.7$ and $\alpha_z=1.4$ (top) and the
radially biased model with $\alpha_{\phi0}=1$, $\alpha_z=1.5$ (bottom).}
\label{fig:1cross}
\end{figure}

\subsection{The velocity ellipsoids}

Fig.~\ref{fig:1srpz} shows the variation within the $(R,z)$ plane of the
velocity dispersions $\sigma_R$, $\sigma_\phi$ and $\sigma_z$ within the
azimuthally biased, non-rotating model with $\alpha_{\phi0}=0.7$ and
$\alpha_z=1.4$. We see that surfaces of constant $\sigma_R$ and $\sigma_\phi$
are quite flattened, while those of $\sigma_z$ are distinctly prolate.
\figref{fig:2srpz} is the analogous figure for the radially biased model with
$\alpha_{\phi0}=1$, $\alpha_z=1.5$. Now the surfaces of constant $\sigma_R$
are decidedly more flattened and those of constant $\sigma_\phi$ are less
flatted, and the surfaces of constant $\sigma_z$ are even more prolate.

\begin{figure}
\centerline{\epsfig{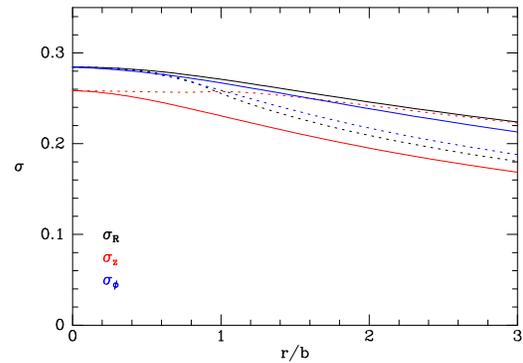}}
\caption{Principal velocity dispersions $\sigma_R$ (black) $\sigma_\phi$
(blue) and $\sigma_z$ (red) along the major (full) and minor (broken) axes of
the model with $\alpha_{\phi0}=1$ and $\alpha_z=1.5$.
 The unit of velocity is $\sqrt{GM/b}$.}\label{fig:rad_sig}
\end{figure}

\figref{fig:1cross} shows the orientation of the velocity ellipsoids of the
tangentially (top) and radially biased non-rotating models. The plots are
remarkably similar. Beyond $r\simeq2b$ the ellipsoids are approximately
aligned with spherical polar coordinates, while at smaller radii the
ellipsoids swing towards alignment with cylindrical polar coordinates. 

\figref{fig:rad_sig} shows how the principal velocity dispersions vary along
the major (full curves) and minor axes of the radially biased model
$\alpha_{\phi0}=1$, $\alpha_z=1.5$. The most remarkable feature is the
extreme flatness of the $\sigma_z$ profile along the minor axis -- there is
no decrease in $\sigma_z$ between the centre and $z=1.25b$. Further out it
converges on the curve for $\sigma_R$ along the major axis, and is
significantly higher than the curve for $\sigma_R$ on the minor axis. The
corresponding plot for the azimuthally biased model $\alpha_{\phi0}=0.7$,
$\alpha_z=1.4$ shows that the curve for $\sigma_z$ remaining flat until it
converges with the curve for $\sigma_R$ along the major axis and follows it
down. Thus beyond the core of any model it seems that $\sigma_z$ varies along
the minor axis much as $\sigma_R$ varies along the major axis. This reflects
the strong connection between these moments and the way the \df\ depends on
$J_r$.

\subsection{Projections of rotating models}

\begin{figure*}
\centerline{\epsfig{file=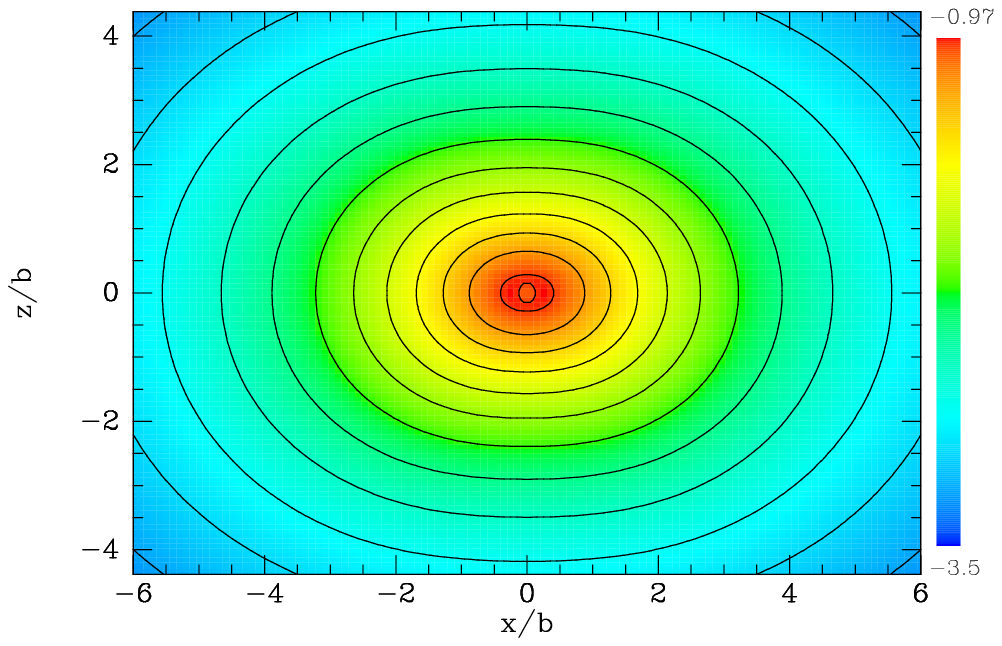,width=.32\hsize}
\epsfig{file=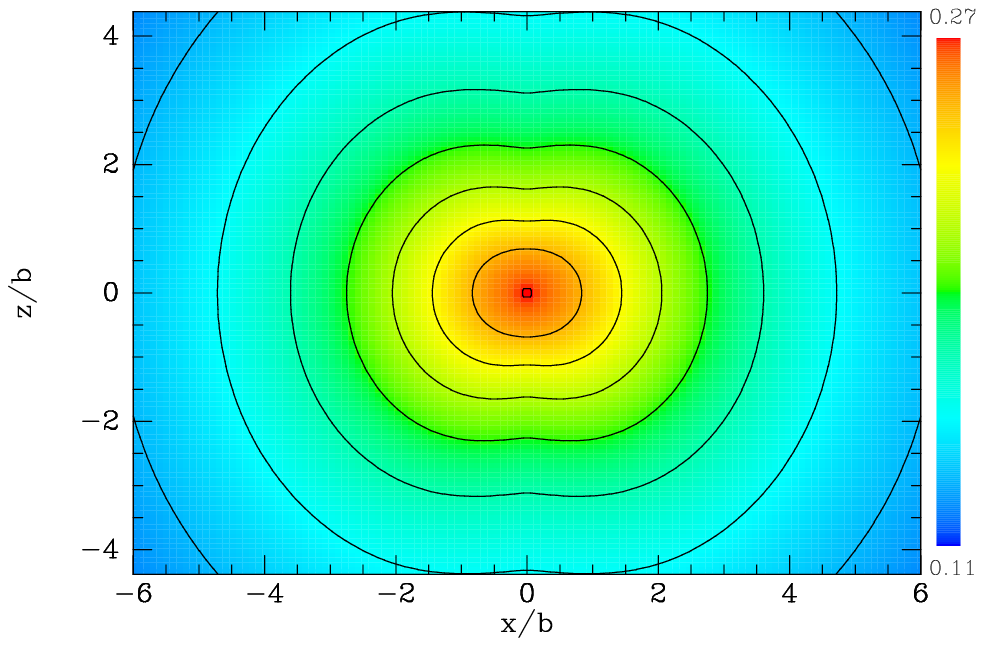,width=.32\hsize}
\epsfig{file=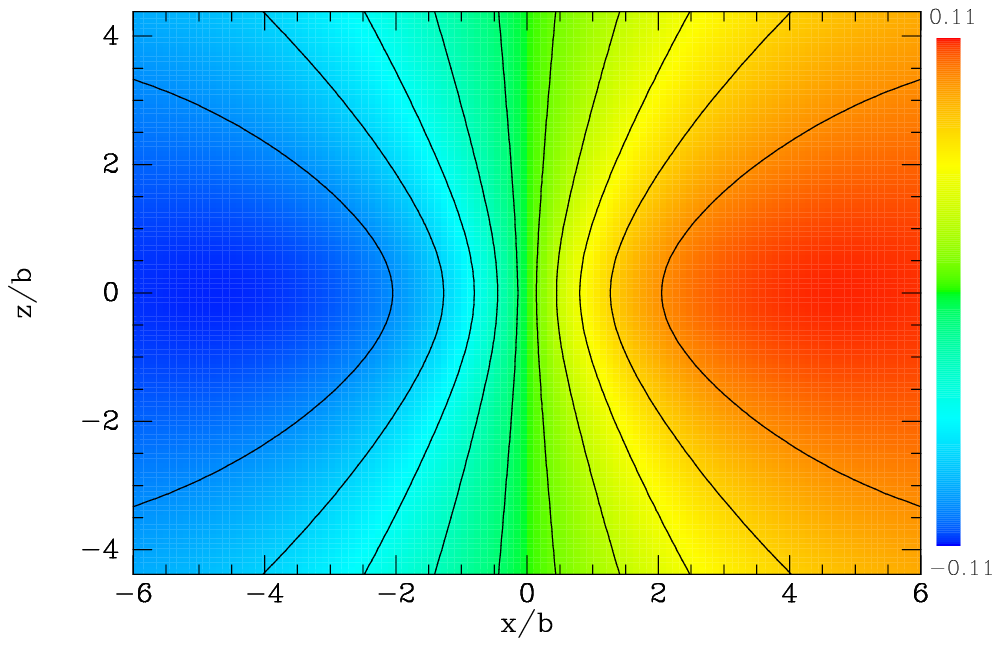,width=.32\hsize}}
\caption{Left: the projected mass density of the model with
$\alpha_{\phi0}=0.7$, $\alpha_z=1.4$ when viewed from the equatorial plane.
Middle and right: the corresponding line-of-sight velocity dispersion and
line-of-sight streaming velocity for the maximally rotating model $k=0.5$.}
\label{fig:sky}
\end{figure*}

The left panel of \figref{fig:sky} shows the projected mass density of the
azimuthally biased model $\alpha_{\phi0}=0.7$, $\alpha_z=1.4$ when it is
viewed from the equatorial plane. The isodensity contours have ellipticity
$\epsilon\simeq0.26$ and are clearly boxy: the disciness coefficient
\citep[][eq.~4.10]{BM98} $a_4/a=-0.013$. Only one of the 48 galaxies in the SAURON
survey presented by \cite{Emsellem} has a more negative disciness. It seems
likely that in real early-type galaxies such a negative disciness from the
spheroid is counteracted by a strongly positive contribution to the disciness
from an embedded disc.

\begin{figure}
\centerline{\epsfig{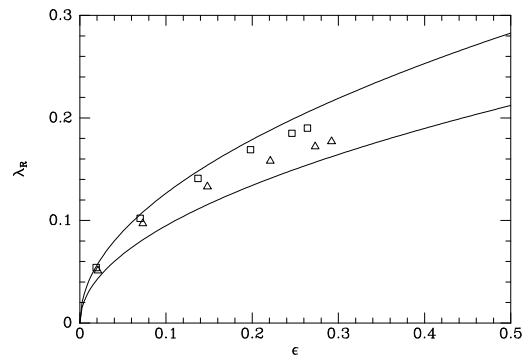}}
\caption{Two maximally rotating models in the rotation-ellipticity plane. 
The average projected rotation parameter defined by equation
(\ref{eq:deflamR}) is plotted against the ellipticity of the isophote at
$R\simeq0.7R_\e$ for inclinations $15^\circ,30^\circ,45^\circ,\ldots$. The
squares are for the azimuthally biased model
$\alpha_{\phi0}=-0.3,\alpha_z=1.4$, while the triangles for the
radially biased model $\alpha_{\phi0}=1,\alpha_z=1.5$. The upper
curve shows the relation $\lamR=0.4\sqrt{\epsilon}$ while the lower curve
shows $\lamR=0.3\sqrt{\epsilon}$. }\label{fig:lambda}
\end{figure}

The centre and right panels of \figref{fig:sky} show the projected velocity
dispersion and mean-streaming velocity of the maximally rotating version of
the model. The contours of constant velocity dispersion are very boxy. The
mean streaming velocity decreases with distance from the major axis, so the
galaxy  does not rotate on cylinders. \cite{Emsellem} defined as a measure of
rotation rate the parameter
\[\label{eq:deflamR}
\lamR={\ex{R\overline{v}}\over\ex{R\sqrt{\sigma^2+\overline{v}^2}}},
\]
 where angle brackets signify luminosity-weighted averages over the part of
the image that lies within the effective radius $R_\e$.  The data plotted yield
$\lamR=0.19$. By reducing the parameter $k$ appearing in equation
(\ref{eq:f_of_k}) we can produce a model with the same ellipticity and any value
of $\lamR$ up to 0.19. By decreasing the inclination at
which we view the model plotted in \figref{fig:sky}, we can construct models
in which $\epsilon$ and $\lamR$ move to the origin on  a certain curve.
In  \figref{fig:lambda} the squares show the points along this curve for
inclinations $15^\circ,30^\circ,45^\circ,\ldots$.

Projection of the radially biased model $\alpha_{\phi0}=1$, $\alpha_z=1.5$
yields isodensity contours of ellipticity $\epsilon=0.29$ and disciness
$a_4/a=-0.015$. The maximally rotating model ($k=0.5$) has $\lamR=0.18$ when
viewed edge-on. The triangles in \figref{fig:lambda} show points in the
$(\epsilon,\lamR)$ plane for  inclinations $15^\circ,30^\circ,45^\circ,\ldots$.

\section{Applications}\label{sec:appls}

\subsection{N-body models}\label{sec:Nbdy}

One of the commonest applications  of self-consistent models that have
analytic distribution functions such as Michie--King models
\citep{Michie,King} and Hernquist models \citep{Hernquist} is the generation
of N-body models that start in an equilibrium configuration rather than
experiencing an early period of violent relaxation towards an uncontrolled
equilibrium. Hence we now explain how the present models can be used to
choose initial conditions for an N-body model.

We start by sampling the isochrone sphere. This is conveniently done by
 defining
\[
u\equiv {v\over\Ve}, \hbox{ where } \Ve(r)\equiv\sqrt{2\Phi(r)}
\]
 is the escape speed, tabulating the integrals
 \[
\rho(r,u)\equiv 4\pi\Ve^3\int_0^u\d u\,u^2f_{\rm
I}\big(\fracj12u^2\Ve^2+\Phi(r)\big)
\]
 on a suitable grid in the rectangle $r\ge0$, $0\le u\le1$. Then the \df\ of
the isochrone sphere can be sampled by picking a number $\xi$ that is
uniformly distributed in [0,1] and finding by interpolation on the grid the
radius $r$ that satisfies $\rho(r,1)=\xi\rho(\infty,1)$ and then choosing a
new random number $\xi$ and determining the value of $u$ that satisfies
$\rho(r,u)=\xi\rho(r,1)$.  Now we evaluate the isochrone's \df\ $f_{\rm
I}(r,v)$ at this radius and kinetic energy.

Next we choose random directions for the position and velocity vectors $\vr$
and $\vv$ and evaluate at the chosen phase-space position the actions
$\vJ(\vr,\vv)$ for the flattened model, and thus evaluate the \df\ $f(\vJ)$.
of this model. We accept this point with probability $kf(\vJ)/f_{\rm
I}(r,v)$, where $k$ is a constant of order unity chosen to ensure that the
ratio never exceeds unity.

\subsection{Stability of models}

For some values of $\valpha$ the model will be unstable to bar formation.
Specifically, if $\alpha_r$ is too small and the model too radially biased,
it will suffer from the radial-orbit instability \citep{FridP,PalmerP}. Similarly, if the
model is set rotating too fast by adding a large odd \df\ $f_-$, it will
develop a rotating bar. Investigation of the onset and development of these
instabilities promises to be a fascinating field of research that would
extend to stellar dynamics the classical work of Dedekind, Jacobi and Riemann
on rotating fluid bodies \citep{Chandra}.

It is likely that for values of $\alpha_r$ that are smaller than some
critical value, $\alpha_{r\,\rm crit}$ a triaxial equilibrium can be found
for the given $f(\vJ)$ that has lower energy than the axisymmetric
equilibrium constructed here, and that axisymmetric models with
$\alpha_r<\alpha_{r\,\rm crit}$ are liable to the radial-orbit instability.
In the limit of infinitely many stars implicit in our discussion, the
transition from an axisymmetric equilibrium to a triaxial one has all the
characteristics of a phase transition. It would be fascinating to know the
nature of this transition.

Currently we are not in a position to construct triaxial models given
$f(\vJ)$ because the St\"ackel Fudge used here seems not to extend to
triaxial potentials. Torus mapping does extend to barred systems, even
rotating ones \citep{KaasalainenB94a,Kaasalainen95b}, so it should be
possible to build models with given $f(\vJ)$ by this method. 

In addition to establishing the relations between axisymmetric and triaxial
equilibria, one would want to follow the dynamics of the instability that
effects the loss of symmetry. The present models are ideal for such an
investigation from two respects. First, the growth of the perturbation can be
followed in angle-action coordinates, which were invented to study the
stability of the solar system and are thus the natural coordinates of
Hamiltonian perturbation theory. Although angle-action coordinates have been
used by a number of authors to study the stability of 
planar, axisymmetric discs \citep[e.g.][]{KalnajsStabil} and  spherical
galaxies \citep[e.g.][]{Weinberg,Saha}, only a small number of rather special
models have been studied in this way for want of a wider range of models for
which angle-action coordinates are available.

Another direction of research into the stability of galaxies that is opened
up by the present models is the method of perturbation particles
\citep{Leeuwin,LeeuwinA}. In this method, invented in unpublished work by G.
Rybicki, the initial model is represented by an analytic \df\ and particles
are used merely to quantify the difference between the model's time-evolving
state and the initial condition. Because the particles start massless and
only gather (positive and negative) mass as the dynamics unfolds, the Poisson
fluctuations in the gravitational potential are much smaller than in a
conventional N-body simulation with the same number of particles, and physics
can be explored to higher precision. To date a major restriction on the use
of the method has been the shortage of dynamically interesting equilibria for
which $f(\vx,\vv)$ is known analytically. Hence our models greatly widen the
range of applicability of this promising method.

\subsection{Modelling early-type galaxies}\label{sec:egals}

The advent of integral-field spectrographs has rejuvenated the study of the
internal dynamics of early-type galaxies \citep{SAURON,Emsellem,Atlas3D}. Now
that it is feasible to quantify the line-of-sight velocity distribution
(LOSVD) over a large part of the image of an E or S0 galaxy, sophisticated
dynamical models can be fitted to the data. In addition to mapping the
variation of the mass-to-light ratio within early-type galaxies, these models
have revealed internal structures in these systems, such as discs and
kinematically decoupled cores. 

The models fitted have been of two types: Schwarzschild orbit-superposition
models \citep{Schwarzschild79,vdV} and models based on the Jeans equations
\citep{Satoh,BDI,CappellariJAM}. In either case the three-dimensional
luminosity distribution is inferred from the photometry under some assumption
of symmetry and inclination angle, a matching potential is adopted, and then
model parameters are adjusted to optimise the fit between predicted and
observed kinematics. 

Schwarzschild modelling is very general but cumbersome because the model
parameters are the weights $w_i$ of some thousands of orbits that together
form a library, and the selection of orbits for the library is an art rather
than a science. In view of these objections, models based on the Jeans
equations are widely used although they lack either generality or rigour
depending on how the modelling is done. The rigorous approach is to assume
that $\sigma_R^2=\sigma_z^2$, which is equivalent to assuming that the \df\
has the restricted ``two-integral'' form $f(E,L_z)$. The \df\ of our own
Galaxy is very different from a two-integral \df\ so it is essential to fit
more general models to observations of external galaxies. 

Recently the ``Jeans Anisotropic Multi-Gaussian Expansion'' (JAM) models of
\cite{CappellariJAM} have been widely used. These models assume that the
principal axes of the velocity ellipsoid are always aligned with the
cylindrical coordinate directions, and that $\sigma_R^2=b\sigma_z^2$, where
$b$ is a constant. Cappellari recognises that the principal axes really align
much more nearly with prolate ellipsoidal coordinates than cylindrical
coordinates (as \figref{fig:1cross} confirms) but argues that on the minor
axis, as in the equatorial plane, the short axis is parallel to the $z$ axis,
while at intermediate latitudes the ellipsoid is nearly spherical. At points
on the minor axis of any of our models the long axis of the velocity
ellipsoid points radially rather than tangentially except in the core. It is
worth understanding why this is so. 

\begin{figure}
\centerline{\epsfig{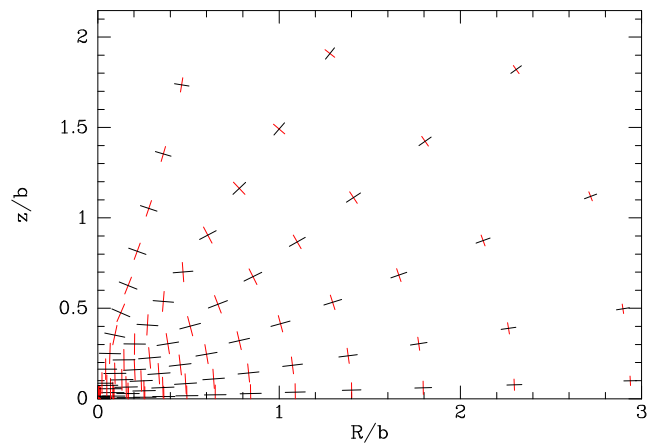}}
\caption{The velocity ellipsoid in a model designed to have tangential bias
along the minor axis.}\label{fig:J1cross}
\end{figure}

Only orbits with small values of $J_\phi$ approach the minor axis, and in
three dimensions an orbit of this type comprises an elliptical annulus that
lies in a plane that is only slightly inclined to the minor axis and
precesses around the axis. If the orbit's radial action is diminished, the
annulus may shrinks within the precessing plane to an elliptical curve, and
in the limit $J_r\to0 $ the orbit becomes a shell orbit. If $J_r$ is
increased, the annulus becomes thick on account of the large radial
excursions along the orbit. At a fixed energy, the sequence of
orbits that starts with the shell orbit and proceeds through orbits of ever
higher eccentricity is a sequence in which $J_r/J_z$ grows from zero to
infinity.  Along the model's minor axis, orbits in the early part of this
sequence stretch the model's velocity ellipsoid in the tangential direction,
while orbits in the later part stretch the ellipsoid radially. Hence the
tangential bias along the minor axis that is assumed in the JAM models
implies dominance by orbits with small $J_r/J_z$. However it is precisely
these orbits that we have suppressed in order to make the model oblate. Hence
there is an essential connection between the flattening of our models and the
radial bias of the velocity ellipsoids along the minor axis beyond the core.
(Inside the core orbits with small $J_r/J_z$ tend to oscillations along the
minor axis rather than thin annuli in a precessing plane and the argument
above does not hold.)

To construct a model in which the velocity ellipsoids behave as assumed when
building a JAM model one might add to one of our models a distinct population
of stars on essentially $J_\phi=0$ shell orbits, i.e. orbits with $J_r/J_z$
and $|J_\phi|/J_z\simeq0$. These orbits are elongated parallel to the model's
symmetry axis, so adding them will reduce the model's flattening. But by the
same token they will tend to dominate the population of stars on the axis and
thus there stretch the velocity ellipsoids tangentially. \figref{fig:J1cross}
shows the structure of the velocity ellipsoids inside a model of this type.
Specifically a new component was  added to the \df\ of the  model
with $\alpha_{\phi0}=0.7$, $\alpha_z=1.4$. The \df\ of the new component is
 \[
f_{\rm I}(\overline{J},\overline{J},\overline{J})
\exp\left[-4(J_r+|J_\phi|)^2/J_z^2\right],
\]
 where $\overline{J}$ is defined by equation (\ref{eq:defsJbar}). With the
new component included, the longest axis of the velocity ellipsoids (plotted
in black) points tangentially out to $\sim2b$ along the minor axis rather
than only within $\sim0.75b$ as in the original model. This change arises
because the velocity ellipsoids do not twist around as one moves outwards
near the minor axis, as they do in the original model. Inevitably, the
addition of the new component diminishes the model's flattening: $\epsilon$
declines from $\sim0.2$ at the half-mass radius to zero at $0.72b$, and the
model is prolate at smaller radii.

It may be that many fast-rotating early-type galaxies do contain a distinct
component that comprises near-polar orbits like the model shown in
\figref{fig:J1cross}, but if this is the case, it is a remarkable
circumstance. In any case the discussion above makes it physically evident that
JAM models are far from generic. If rigorous construction of models with
their presumed properties is possible, a prerequisite would seem to be a \df\
that has at least two peaks on each surface of constant energy in action
space: most stars must be associated with a peak in the region of low $J_z$
and be responsible for the model's flattening, while a second peak near the
$J_r=J_\phi=0$ vertex of that surface is responsible for the tangential
orientation of the velocity ellipsoids along the model's minor axis.

We have concentrated on exceptionally simple \df s. Galaxies with different
shapes and kinematics could be constructed using \df s that are either linear
combinations of the \df s explored here, or involve other simple functional
forms for $f(\vJ)$. In particular it would be straightforward to make a model
that, like NGC 4550, has counter-rotating discs \citep{Rubin}, or a galaxy
that has a kinematically decoupled core. 

\subsection{Multi-component galaxies}

All galaxies are thought to contain substantial amounts of dark matter, and
most galaxies contain stellar discs in addition to a spheroidal stellar
component. It is straightforward to generalise the present models to
multi-component systems: one simply chooses a \df\ $f_i(\vJ)$ for each
component. A very convenient aspect of this choice is that  each
component's mass $m_i$ is specified by the integral $\int\d^3\vJ\,f_i(\vJ)$, so it is
determined before one solves for the model's spatial form. The latter is done
just as in Section \ref{sec:relax} with the \df\ given by $\sum_i f_i$.
Naturally the observables of the relaxed model are obtained by integrating
only over the \df s of the stellar components. We hope shortly to present
models of our Galaxy that have been constructed in this way by combining a
\df\ for of the type described by \cite{Binney12b} with a \df\ for the dark
halo of the type described by \cite{Pontzen}.

\section{Conclusions}\label{sec:conclude}

We have presented a new type of self-consistent model of hot, axisymmetric
stellar systems that have specified \df s $f(\vJ)$, where $\vJ$ is the triple
of action values. The even part of the \df\ is specified by two dimensionless
parameters $\alpha_{\phi0}$ and $\alpha_z$. When $\alpha_z>1$ the model
becomes oblate. If $\alpha_{\phi0}\simeq1$ the model is radially biased,
while dropping $\alpha_{\phi0}$ below unity reduces the radial bias and, for
sufficiently small values of $\alpha_{\phi0}$, the model becomes azimuthally
biased. The odd part of the \df, which does not contribute to the model's
density profile $\rho(\vx)$, is controlled by two dimensionless parameters,
$\chi\ge0$, which controls the steepness of the rotation curve at the centre,
and $0\le k\le0.5$, which determines how fast the model rotates: increasing
$k$ both speeds up the model's rotation and diminishes the magnitude of the
azimuthal velocity dispersion $\sigma_\phi$.

We focused on the observables of just two exemplary models, a tangentially
biased model with $\alpha_{\phi0}=0.7$ and a radially biased model with
$\alpha_{\phi0}=1$.  Both models achieve peak ellipticities
$\epsilon\simeq0.3$ at about $R_\e/2$ and are distinctly boxy when seen
edge-on.  In each model the velocity ellipsoids are aligned with cylindrical
coordinates within the core, but beyond the core they quickly align with
radial polar coordinates, and along the minor axis the ellipsoids point
towards the centre rather than tangentially. We have argued that the only way
to make the velocity ellipsoids point tangentially at significant distances
down the minor axis is to include a distinct component of stars on nearly
polar orbits.

The models could provide initial conditions for N-body models that start from
perfect equilibrium, something that is possible only when the \df\ is explicitly
known. Since the angle-action coordinates of any point in the phase space of
one of these models are readily computed, the models provide perfect testbeds
for studies of galactic stability and evolution.

The present models differ from Schwarzschild models in having vastly fewer
free parameters: four versus the number of orbits in the orbit library
employed. They differ from models based on the Jeans equations in providing
full velocity distributions rather than just the first two moments of the
distributions. It would be straightforward to extend these models in various
directions. For example, it is easy to devise other approaches than simple
action scaling to move from the \df\ $f(H)$ of an ergodic model to a
three-integral \df\ f(\vJ), and we are currently exploring one of these
alternatives.  Another direction in which the present work is being extended
is to multi-component systems, in which stars and dark matter have distinct
distribution functions, and, in the case of our Galaxy, the thin disc, thick
disc and halo stellar populations all have distinct \df s.

We started from the isochrone sphere because it provides an analytic
expression for $H(\vJ)$. In a forthcoming paper  we will show how to obtain
a good approximation to an analogous expression for any spherical model, and
thus extend the present work to other popular spherical systems, such as the
\cite{Hernquist} sphere.

\section*{Appendix I: Improving the evaluation of actions}\label{sec:uvapp}

The scheme for the evaluation of actions is that described in B12a except for
modifications described here.

\subsection*{A redefined third integral}

A redefinition of the third integral extracted from the St\"ackel Fudge
proves expedient: in the notation of B12a we now use as the third integral 
\begin{eqnarray}\label{eq:defsEr}
E_r&=&{p_u^2\over2\Delta^2\cosh^2u_0}
+{L_z^2\over2\Delta^2\cosh^2u_0}\left(\sinh^{-2}u-\sinh^{-2}u_0\right)\nonumber\\
&&+{\delta
U(u)\over\cosh^2u_0}-{E\over\cosh^2u_0}\left(\sinh^2u-\sinh^2u_0\right),
\end{eqnarray}
 where $u_0$ is the location of the minimum in the effective potential
$\delta U(u)$ that governs oscillations in $u$. Thus defined $E_r$ is related to the third integral $I_3$ defined by
equation (9) of B12a by
\[
E_r=-{I_3(p_u,u)-I_3(0,u_0)\over\cosh^2u_0}.
\]
 Subtracting $I_3(0,u_0)$ ensures that $E_r=0$ for a shell orbit, and
dividing by $\cosh^2u_0$ ensures that $E_r$ is almost invariant under a change in
$\Delta$. In fact, $E_r$ has many of the properties one expects of the ``radial
energy'' and is thus a nicely physical third integral.

\begin{figure}
\centerline{\psfig{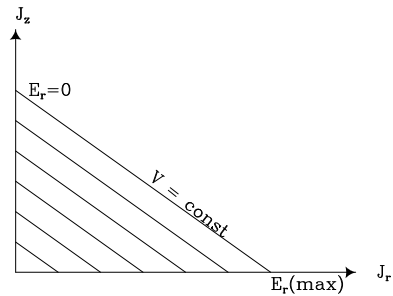}}
\caption{A typical triangle within the  orbit grid. The sloping lines are
ones of constant $\Enc$.}\label{fig:triangle}
\end{figure}

\subsection*{Interpolation}

As discussed by B12a, the integrals over velocity space are significantly
accelerated if actions are obtained by interpolation from variables that can
be computed from $(\vx,\vv)$ algebraically rather than by evaluation of
integrals over $u$ and $v$. Interpolation errors prove to dominate the error
budget, but they can be minimised by judicious choice of the grid.
Three suitable algebraic quantities are the difference $\Enc=E-E_{c}$ between
the orbit's energy $E$ and the energy $E_c(L_z)$ of the circular orbit with
its value of $L_z$, and the radial energy $E_r$ (eq.~\ref{eq:defsEr}).  Together
$(L_z,\Enc,E_r)$ specify the effective potentials in which the star is
presumed to oscillate in $u$ and $v$, and thus make the actions $J_r=J_u$ and
$J_z=J_v$ available by quadrature. Our approach is to tabulate the values of
these integrals on a grid in $(L_z,\Enc,E_r)$ 

It is impracticable for the grid to cover the whole of action space, which is
infinite.  Good accuracy can be achieved, however, by covering the part of
action space on which the \df\ is largest. This part occupies the tetrahedron
that lies between the approximately planar surface $H(\vJ)=E_{\rm max}$ and
the origin (\figref{fig:Jspace}). We take the primary axis to be the $L_z$
axis, which in \figref{fig:Jspace} runs left to right. At each grid point on
this axis we have to consider a triangular domain. At low $L_z$ the triangles
are large, and they shrink with increasing $L_z$. \figref{fig:triangle} shows
one of these triangles. Running across it is a series of lines of constant
$\Enc$.

The grid points $(L_z,\Enc,E_r)$ with known actions are defined as follows.
For each grid value of $L_z$ we choose a grid of values of $\Enc$ by taking
equal increments in the speed $V$ that determines $\Enc=\fracj12V^2$.  Next
we find the intersection with the equatorial plane of the shell orbit
($J_r=0$) of the given values $(L_z,\Enc)$ by locating the minimum of the
effective potential $\delta U(u)$. Once this point has been located, we can
compute the speed with which these orbits pass through the point and we
compute the values of $E_r$ and the actions $(J_r,J_z)$ for a series of
orbits that pass through the point moving at equally spaced angles $\psi$
with respect to the vertical: when $\psi=0$, the shell orbit is generated,
while when $\psi=\pi/2$, a planar orbit is generated, which has the largest
value of $J_r$ of any orbit with the given $(L_z,\Enc)$. In
\figref{fig:triangle} the generated orbits form a sequence of points that
runs along one of the lines of constant $\Enc$ that slope down from the
vertical axis to the horizontal axis.

Interpolation within this grid works as follows. Given a triple
$(L_z,E,E_r)$, we determine between which two planes of constant $L_z$ the
triple lies; let $\d x$ be the fractional displacement of the triple from the
lower plane. Then by linear interpolation on values stored for each grid
value of $L_z$ we estimate $E_c$ for the given angular momentum and thus
obtain $\Enc=E-E_c$ and thus the speed $V=\sqrt{2\Enc}$.  We interpolate
linearly between the maximum speeds $V_{\rm max}$ used for the adjacent
angular-momentum grid points to produce the renormalised variable $s=V/V_{\rm
max}(L_z)$. 

The adjacent grid planes of constant $L_z$ have grid points along lines of
constant $s$, and we find between which two such lines our point lies on both
planes. In a given plane, let $\d y$ denote the fractional displacement
implied by $s$ above the line of smaller $s$. 

Now that we have identified the nearest grid lines $s=\hbox{constant}$, we
have to identify the points on these lines that most nearly correspond to our
point. By linear interpolation on the end points of our lines we estimate the
maximum value of $E_r$ along the (non-grid) line of constant $s$ through our
point, and use this to produce the dimensionless coordinate $h=E_r/E_r({\rm
max})$. Then along each grid line we find the actions associated with this value
of $h$, and finally combine these four values of each of $J_r$ and $J_z$ using
the weights provided by $\d x$ and $\d y$.

A tiresome complication is that for $s=0$ the line $s=\hbox{constant}$
degenerates into the single point $J_r=J_z=0$, and these actions have not
been explicitly calculated. So, the lowest value of $s$ in the stored values
for a plane is greater than zero, and our value of $s$ may prove to be
smaller than this. In this case we have to interpolate between zero actions
and the actions on the line associated with the first stored line
$s=\hbox{constant}$.

The array structure used to store the actions and values of $E_r({\rm max})$
are non-cubical: the number of values of $V$ stored decreases with increasing
$L_z$ from $\simeq100$ to $\simeq10$. At fixed $L_z$ the number of values of
$\sqrt{E_r}$ stored for each value of $\Enc$ increases from $\simeq100$
to $\simeq10$. The stored values of $V$ increase from a small value to
$0.98(-2\Phi(R_c,0))^{1/2}$.

The $L_z$ grid has uniform increments between a value as close as possible to
zero (there is a coordinate singularity at zero, so we must avoid it) and a
largest value of $L_z=0.98R_{\rm max}(-2\Phi(R_{\rm max},0))^{1/2}$, where
$R_{\rm max}$ is the largest radius at which moments are required.

\subsection*{Choosing the focal distance $\Delta$}

B12a shows that for orbits in the extended solar neighbourhood
$\Delta\simeq3.5\kpc$ was a reasonable choice for the distance down the $z$
axis of the focus of the confocal $(u,v)$ coordinate system upon which the
St\"ackel Fudge relies. B12a shows also that results are not
sensitive to this parameter. Now that we need to use the St\"ackel fudge at
all points of a variety of potentials, one  needs an algorithm for choosing
$\Delta$ for any orbit in any potential.

Since a shell orbit $J_r=0$ should lie on a surface $u=\hbox{const}$, a
natural procedure is to compute shell orbits $J_r=0$ for relatively large
$J_z$, fit them with ellipses in the $(R,z)$ plane, and to read off
$\Delta^2$ from these ellipses. 

There is a shell orbit that reaches to the $z$ axis for any pair $(E,L_z)$
and for the $q=0.7$ flattened isochrone one can fit ellipses to these
orbits over the entire $(E,L_z)$ grid. One finds that when the corresponding
curves of $\Delta^2(E)$ at fixed $L_z$ for different values of $L_z$ are
plotted together, they lie one on top of another. Consequently, we take
$\Delta$ to be a function of $E$ alone. For a given, small value of $J_z$ we
compute the maximal shell orbits on the grid in $E$ and fit them with ellipses.

From the orbit integrations we have the point $(R_0,0)$ in the middle of the
orbit $J_r=0$. We start by seeking an ellipse that passes through this point
and near to some other point  $(R,z)$ on the orbit. An ellipse through
$(R_0,0)$ is
\[
R=R_0\sin v\quad z=\sqrt{R_0^2+\Delta^2}\cos v
\]
and we minimise
 \[
s^2=(R_0\sin v-R)^2+(\sqrt{R_0^2+\Delta^2}\cos v-z)^2
\]
 with respect to $v$ at fixed $\Delta$ by seeking the solution to
\begin{eqnarray}\label{eq:dsdv}
0&=&\fracj12{\pa s^2\over\pa v}=(R_0\sin v-R)R_0\cos v\nonumber\\
&&\qquad-(\sqrt{R_0^2+\Delta^2}\cos v-z)\sqrt{R_0^2+\Delta^2}\sin v\\
&=&z\sqrt{R_0^2+\Delta^2}\sin v-RR_0\cos v-\Delta^2\cos v\sin v.\nonumber
\end{eqnarray}
 From this we obtain $\pa v/\pa\Delta^2$:
\begin{eqnarray}
0&=&{z\over2\sqrt{R_0^2+\Delta^2}}\sin v-\fracj12\sin2v\nonumber\\
&&+\Bigl(z\sqrt{R_0^2+\Delta^2}\cos v
+RR_0\sin v-\Delta^2\cos2v\Bigr){\pa v\over\pa\Delta^2}
\nonumber
\end{eqnarray}
so
\[
{\pa v\over\pa\Delta^2}={{z\over\sqrt{R_0^2+\Delta^2}}\sin
v-\sin2v\over2( z\sqrt{R_0^2+\Delta^2}\cos v
+RR_0\sin v-\Delta^2\cos2v)}
\]
 Finally we have
 \begin{eqnarray}\label{eq:dsdD}
0&=&{\pa s^2\over\pa\Delta^2}={\pa s^2\over\pa v}{\pa
v\over\pa\Delta^2}+2(\sqrt{R_0^2+\Delta^2}\cos v-z){\cos
v\over2\sqrt{R_0^2+\Delta^2}}\nonumber\\
&=&{\pa s^2\over\pa v}{\pa
v\over\pa\Delta^2}
+\left(\cos v-{z\over\sqrt{R_0^2+\Delta^2}}\right)\cos v.
\end{eqnarray}
 The code uses Brent's algorithm to solve (\ref{eq:dsdv}) for $v$ and then
 (\ref{eq:dsdD}) for $\Delta^2$.

\section*{Appendix II: Computing $\Phi$}\label{sec:Phiapp}

The grid on which the density is computed is based on Gauss-Legendre
integration with respect to colatitude $\theta$ and finite-difference
integration in spherical radius $r$. The potential is obtained as a sum
 \[
\Phi(r,\theta)=\sum_{l=0,2,\ldots}^N\phi_l(r)P_l(\cos\theta)
\]
 over even-order Legendre polynomials, with the coefficients $\phi_l(a)$ obtained by
interpolation. We take $N=8$. The radial grid points are
 \[
r_i=a_0\sinh(i\delta)\quad i=0,\ldots,N_r-1,
\] 
 where
\[
\delta={1\over N_r-1}\hbox{asinh}(r_{\rm max}/a_0).
\]
 Consequently, the grid points are uniformly spaced in $r$ for $r\ll a_0$ and uniformly
spaced in $\ln r$ for $r\gg a_0$.

To evaluate the $\phi_l$ one has to compute integrals (cf eq.~2.95 of BT08)
 \[
I^{(a)}_l\equiv\int_0^r\d a\,a^{l+2}\rho_l(a)\hbox{ and }
I^{(b)}_l\equiv\int_r^\infty {\d a \over a^{l-1}}\rho_l(a),
\]
 where $\rho_l(a)$ is obtained by integrating $\rho(a,\theta)P_l(\cos\theta)$
with respect to $\cos\theta$.

 At small $a$ we know that $\rho_l(a)\to0$ like $a^l$, so we estimate the
integrals between grid points at small $a$ by
 \[
\int_{r_i}^{r_{i+1}}\d a\,a^{l+2}\rho_l(a)\simeq
\fracj12\left({\rho_l(r_{i+1})\over r_{i+1}^l}+{\rho_l(r_i)\over
r_i^l}\right){r_{i+1}^{2l+3}-r_{i}^{2l+3}\over 2l+3}
\]
and
\[
\int_{r_i}^{r_{i+1}} {\d a \over a^{l-1}}\rho_l(a)\simeq
\fracj14\left({\rho_l(r_{i+1})\over r_{i+1}^l}+{\rho_l(r_i)\over
r_i^l}\right)({r_{i+1}^2-r_{i}^2}).
\]
 Beyond a fiducial radius $a_c$ these integrals are estimated as
 \[
\int_{r_i}^{r_{i+1}}\d a\,a^{l+2}\rho_l(a)\simeq
\fracj12\left[\rho_l(r_{i+1})+\rho_l(r_i)\right]{r_{i+1}^{l+3}-r_{i}^{l+3}\over l+3}
\]
and
\[
\int_{r_i}^{r_{i+1}} {\d a \over a^{l-1}}\rho_l(a)\simeq
\fracj12\left[\rho_l(r_{i+1})+\rho_l(r_i)\right]{{r_{i+1}^{2-l}-r_{i}^{2-l}\over
2-l}}
\]
 with appropriate special treatment of the case $l=2$. The radial derivatives of $\Phi$ are  obtained at the grid points by analytic
differentiation of the power-series expansion:
 \begin{eqnarray}
{\pa\Phi\over\pa r}&=&-4\pi G\sum_l P_l(\cos\theta)\left(-{l+1\over
r^{l+2}}I^{(a)}_l +lI^{(b)}_lr^{l-1}\right)\nonumber\\
{\pa^2\Phi\over\pa r^2}&=&-4\pi G\sum_l P_l(\cos\theta)\biggl({(l+2)(l+1)\over
r^{l+3}}I^{(a)}_l\\
&&+l(l-1)r^{l-2}I^{(b)}_l-(2l+1)\rho_l(r)\biggr).\nonumber
\end{eqnarray}
 Values of the $\phi_l(r)$ and
their first two radial derivatives are obtained at general points by
interpolation from the grid-point values. Since the first non-trivial term in
the power-series expansion of $\Phi$ around the origin is $\propto r^2$,
quadratic interpolation in $r$ between the nearest three grid points is used
inside the tenth radial grid point. Linear interpolation is used further out.

When the potential is required at a radius $r>r_{\rm max}$ that lies outside
the grid, it is readily obtained from the potential at the edge of the grid
because in the vacuum $\phi_l(r)\propto r^{l+1}$.

The tangential derivatives of $\Phi$ are obtained by analytic
differentiation of the Legendre polynomials. 

\label{lastpage}

\end{document}